\documentclass[sigplan,nonacm]{acmart}

\usepackage{natbib}

\usepackage{amsmath,amssymb,amsfonts,amsthm}
\usepackage{graphicx}
\usepackage{xcolor}
\usepackage{microtype}
\usepackage[italic]{mathastext}
\usepackage{microtype}
\usepackage{mathastext}
\usepackage{libertine}
\usepackage[T1]{fontenc}
\usepackage{textcomp}
\usepackage{zi4}
\usepackage[all]{nowidow}
\usepackage{flushend}
\usepackage{fancyhdr}
\usepackage{url}
\usepackage{listings}

\usepackage{makecell}

\usepackage{algorithmic}
\theoremstyle{definition}

\usepackage[linesnumbered]{algorithm2e}
\usepackage{mathtools}
\usepackage{booktabs}
\usepackage{adjustbox}

\usepackage{newfloat}

\usepackage[small,compact]{titlesec}
\usepackage{caption}
\usepackage{color}
\usepackage[normalem]{ulem}
\usepackage{xspace}
\usepackage{booktabs} % for professional tables

\usepackage{makecell}
\usepackage{multirow}
\usepackage{comment}
\usepackage{hhline}
\usepackage[para]{threeparttable} 
\makeatletter
\def\TPT@doparanotes{\par
   \prevdepth\z@ \TPT@hsize
   \TPTnoteSettings
   \parindent\z@ \pretolerance 8
   \linepenalty 200
   \renewcommand\item[1][]{\relax\ifhmode \begingroup
       \unskip
       \advance\hsize 10em % \hsize is scratch register, based on real hsize
       \penalty -45 \hskip\z@\@plus\hsize \penalty-19
       \hskip .07\hsize \penalty 9999 \hskip-.15\hsize
       \hskip .01\hsize\@plus-\hsize\@minus.01\hsize 
       \hskip 6em\@plus .3em
      \endgroup\fi
      \tnote{##1}\,\ignorespaces}%
   \let\TPToverlap\relax
   \def\endtablenotes{\par}%
}
\makeatother

\usepackage{hyperref}

\RequirePackage[nameinlink]{cleveref} % 'nameinlink' to mimic \autoref's behavior
\crefname{chapter}{Chapter}{Chapters}
\crefname{section}{Section}{Sections}
\crefname{subsection}{Subsection}{Subsections}
\crefname{equation}{Equation}{Equations}
\crefname{definition}{Definition}{Definitions}
\crefname{assumption}{Assumption}{Assumptions}
\crefname{theorem}{Theorem}{Theorems}
\crefname{figure}{Figure}{Figures}
\crefname{table}{Table}{Tables}
\crefname{BOX}{Box}{Boxes}
\crefname{algorithm}{Algorithm}{Algorithm}

\let\autoref\cref % set \autoref as an alias for \cref

\DeclareMathOperator*{\argmin}{arg\,min}

\newcommand{\insertFigure}[2]{
  \begingroup%
    \edef\tempfilename{figures/#1.pdf}%
    \begin{figure}[t]
        \centering
        \includegraphics[width=\linewidth]{\tempfilename}
        \vspace{-7mm}
        \caption{\small #2}
        \vspace{-3mm}
        \label{fig:#1}
    \end{figure}
  \endgroup%
}

\newcommand{\insertWideFigure}[2]{
  \begingroup%
    \edef\tempfilename{figures/#1.pdf}%
    \begin{figure*}[h]
        \centering
        \includegraphics[width=\textwidth]{\tempfilename}
        \vspace{-7mm}
        \caption{\small #2}
        \vspace{-4mm}
        \label{fig:#1}
    \end{figure*}
  \endgroup%
}

\DeclareFloatingEnvironment[
  fileext=lob,
  listname={List of Boxes},
  name=Box,
  placement=htp,
]{BOX}

\ifx\forsubmission\undefined
\newcommand{\TODO}[1]{\textcolor{red}{TODO: #1}}
\newcommand{\FT}[1]{\textcolor{orange}{FT: #1}}
\newcommand{\HK}[1]{\textcolor{blue}{HK: #1}}

\else
\newcommand{\TODO}[1]{\textcolor{red}{}}
\newcommand{\CM}[1]{\textcolor{magenta}{}}
\newcommand{\FT}[1]{\textcolor{orange}{}}
\newcommand{\HK}[1]{\textcolor{blue}{}}
\newcommand{\MP}[1]{\textcolor{teal}{}}

\fi

\newcommand{\squishlist}{
 \begin{list}{$\bullet$}
  { \setlength{\itemsep}{0pt}
     \setlength{\parsep}{3pt}
     \setlength{\topsep}{3pt}
     \setlength{\partopsep}{0pt}
     \setlength{\leftmargin}{1.5em}
     \setlength{\labelwidth}{1em}
     \setlength{\labelsep}{0.5em} } }

\newcommand{\squishlisttwo}{
 \begin{list}{$\bullet$}
  { \setlength{\itemsep}{0pt}
     \setlength{\parsep}{0pt}
    \setlength{\topsep}{0pt}
    \setlength{\partopsep}{0pt}
    \setlength{\leftmargin}{2em}
    \setlength{\labelwidth}{1.5em}
    \setlength{\labelsep}{0.5em} } }

\newcommand{\squishend}{
  \end{list}  }

\newcommand{\betterparagraph}[1]{\noindent \textbf{#1. }}

\newcommand{\norm}[1]{\left\lVert#1\right\rVert}

\crefname{algorithm}{Algorithm}{Algorithms}
\crefname{listing}{Code}{Code}

\definecolor{codegreen}{rgb}{0,0.6,0}
\definecolor{codegray}{rgb}{0.5,0.5,0.5}
\definecolor{codepurple}{rgb}{0.58,0,0.82}
\definecolor{backcolour}{rgb}{0.95,0.95,0.92}

\lstdefinestyle{mystyle}{
    backgroundcolor=\color{backcolour},   
    commentstyle=\color{codegreen},
    keywordstyle=\color{magenta},
    numberstyle=\tiny\color{codegray},
    stringstyle=\color{codepurple},
    breakatwhitespace=false,         
    breaklines=true,                 
    captionpos=b,                    
    keepspaces=true,                 
    numbers=left,                    
    numbersep=5pt,                  
    showspaces=false,                
    showstringspaces=false,
    showtabs=false,                  
    tabsize=2,
    basicstyle=\tiny
}

\newcommand{\accelerator}{\textsc{{D-com}}\xspace}

\begin{document}
%%%%%%%%%%%%%%%%%%%%%%%%%%%%%%%%%%%%%%%%
%%%%%%%%%%%%%% -- UPDATE -- %%%%%%%%%%%%%%%
%\newcommand{\hpcasubmissionnumber}{NaN}
\title{\accelerator: Accelerating Iterative Processing to Enable Low-rank Decomposition of Activations}
%%%%%%%%%%%%%%%%%%%%%%%%%%%%%%%%%%%%%%%%

\author{Faraz Tahmasebi}
\affiliation{%
  \institution{University of California, Irvine}
  \department{Electrical Engineering and Computer Science}
  \city{Irvine}
  \state{CA}
  \country{USA}
}
\email{tahmasef@uci.edu}

\author{Michael Pelluer}
\affiliation{%
  \institution{NVIDIA}
  \department{}
  \city{Westford}
  \state{MA}
  \country{USA}
}
\email{mpellauer@nvidia.com}

\author{Hyoukjun Kwon}
\affiliation{%
  \institution{University of California, Irvine}
  \department{Electrical Engineering and Computer Science}
  \city{Irvine}
  \state{CA}
  \country{USA}
}
\email{hyoukjun.kwon@uci.edu}

%%%%%%%%%%%%%%%%%%%%%%%%%%%%%%%%%%%%%%%%
%%%%%%%% -- ONLY FOR CAMERA READY -- %%%%%%%%
%\def\hpcacameraready{} % Uncomment to build camera-ready version
%\newcommand{\hpcapubid}{0000--0000/00\$00.00}
%\newcommand\hpcaauthors{First Author$\dagger$ and Second Author$\ddagger$}
%\newcommand\hpcaaffiliation{First Affiliation$\dagger$, Second Affiliation$\ddagger$}
%\newcommand\hpcaemail{Email(s)}

%%%%% -- ARTEFACT EVALUATION RESULTS -- %%%%%%
% Uncomment the following based on the badges that were awarded to this paper
%\def\aeopen{}           % The artifact is publically available
%\def\aereviewed{}     % The artefact has been reviewed
%\def\aereproduced{} % The results have been reproduced
%%%%%%%%%%%%%%%%%%%%%%%%%%%%%%%%%%%%%%%%

%\input{hpca-template}

%%%%%%%%%%%%%%%%%%%%%%%%%%%%%%%%%%%%%%%%
%%%%%%%% -- PAPER CONTENT STARTS -- %%%%%%%%%

\begin{abstract}

The computation and memory costs of large language models kept increasing over last decade, which reached over the scale of 1T parameters.
To address the challenges from the large scale models, model compression techniques such as low-rank decomposition have been explored.
Previous model decomposition works have focused on weight decomposition to avoid costly runtime decomposition, whose latency often significantly exceeds the benefits from decomposition (e.g., 38\% more end-to-end latency when running Llama2-7b on A100 with 4K sequence length with activation decomposition compared to no decomposition).

In this work, we debunk such observations and report that the input decomposition can be significantly beneficial with a proper choice of decomposition algorithm and hardware support.
We adopt progressive decomposition algorithm, Lanczos algorithm, and design a co-accelerator architecture for the decomposition algorithm.
To address the memory-boundness of the decomposition operation, we introduce a novel compute replication methodology that moves the operation toward compute-bound region, which enables 6.2$\times$ speedup in our evaluation.
We also develop an output shape-preserving computation scheme that eliminates decomposition costs in consecutive layers.
To compensate model quality loss from compression, we introduce a multi-track decomposition approach that separately handles outlier channels for high accuracy and low perplexity with minimal computational costs.
Combined together, our accelerator, \accelerator,  provides 22\% end-to-end latency improvements compared to A100 GPU at the cost of small model quality degradation (e.g., 3\% on AI2 Reasoning Challenge task).

\end{abstract}

\maketitle
\section{Introduction}
\label{sec:intro}

In recent years, the scale of large language models (LLMs) have dramatically increased, both in terms of parameter count and the length of input sequences they are expected to process. Models such as GPT-3 (175B parameters)~\cite{few_shots}, PaLM (540B)~\cite{PaLM}, and GPT-4 (estimated \>1T parameters, according to industry reports~\cite{GPT-4}), exemplify this trend. Simultaneously, the context window — the maximum number of tokens a model can attend to — has also expanded rapidly. GPT-2 supported 1,024 tokens, while GPT-3 extended this to 2,048. More recent models like Claude 2 and GPT-4-turbo support up to 100,000 and 128,000 tokens respectively~\cite{GPT-4,claude2}, allowing them to process entire books or long documents in a single pass. This exponential growth in model size and context length comes with significant memory and compute costs, especially during inference and training on long sequences. As a result, optimizing the internal representations, particularly activations, is becoming critical for scaling LLMs efficiently.

To address the computational and memory challenges posed by large-scale LLMs, several model compression and acceleration techniques have been actively explored. Quantization reduces the precision of model weights and activations, enabling faster computation and reduced memory usage with minimal impact on accuracy~\cite{gptq}. Pruning methods remove redundant weights or entire neurons based on importance metrics, often yielding sparse networks with lower compute requirements~\cite{pruning}. Knowledge distillation compresses a large “teacher” model into a smaller “student” model by transferring output behavior or intermediate representations~\cite{knowledge_distillation}. While these approaches primarily target model weights, low-rank decomposition provides a complementary strategy focused on reducing the dimensionality of activations and weight matrices by exploiting their linear structure. Specifically, decomposition techniques such as SVD or Tucker decomposition can approximate high-dimensional tensors with fewer parameters, offering a promising route to lower runtime and memory without retraining the model. 

Applying Low-Rank Decomposition on the model has been investigated in previous works to some extent. LoRA\cite{LoRA} exploits a low-rank auxilary matrix to train for fine-tuning and eventually add it to the main weight matrix. TIE framework~\cite{tie} introduces an inference-efficient approach for deep neural networks that are compressed using Tensor Train decomposition.
Saha, R, et. al in~\cite{low_prec_low_rank} proposes a combination of low-precision matrix and low-rank high precision matrix to approximate the weights of the model.. All previous works have been focusing on applying low-rank decomposition on the model, while activation decomposition hasn't been explored yet. 
In this work, we enable the low-rank decomposition of activations as well as weights. In order to mitigate the model quality degradation, we reinforce outlier extraction for input to separate out the crucial activations. Doing so, we reduce the inference memory footprint and computation of a layer significantly, resulting $22\%$ end-to-end model runtime while keeping the model quality high. Based on our observed pattern, outlier extraction is done channel-wise to keep the memory footprint and computation overhead small. 
However, model decomposition can be done offline, but input decomposition must happen real-time, diminishing the decomposition benefits. Thus, we also propose a decomposer accelerator that optimizes the associated iterative computations in decomposition.~\autoref{fig:Overview} depicts an overview of our work.

\insertFigure{Overview}{An overview of~\accelerator.}

Our main contributions are as follows.
\begin{itemize}
    {\item We enable input decomposition as a novel approach to reduce the LLM inference memory footprint and computation. We also explore the decomposition of both inputs and the model and its effect on the model quality to push the computation optimization further.}
    {\item Inspired by~\cite{IISWC_paper}, we search for a good trade-off between model quality and inference runtime.}
    {\item We reinforce channel-wise outlier extraction to eliminate the negative effect of low-rank approximation. Extracting $\leq$ 5\% of activations can significantly improve the model quality after decomposition. The channel-wise granularity of this approach keeps memory footprint and computation overhead relatively small.}
    {\item We propose \accelerator to be deployed alongside GEMM accelerators to minimize the input decomposition in the computational graph. We expand the computations of iterative processes to increase utilization of compute and memory bandwidth resources . We achieve $3.8 \times$ speedup over a single non-decomposed layer and $8.74 \times$ speedup over a single decomposed layer on 4 A100 GPUs~\cite{a100}.}
\end{itemize}
In the next section, we will discuss background knowledge about Language models and low-rank decomposition algorithms.
\section{Background and Motivation}
\label{sec:background}
\subsection{Language Model's Compute Graph}

At the core of most language models lies the Transformer architecture, composed of a stack of identical layers that define the model’s compute graph. Each Transformer layer processes a sequence of hidden states through a series of structured operations. As illustrated in Figure X, the Transformer layer processes an input tensor \(X \in \mathbb{R}^{S \times H}\)
where $S$ is the sequence length and $E$ is the embedding dimension. The input first passes through a multi-head self-attention block, where it is linearly projected into queries, keys, and values, followed by scaled dot-product attention and an output projection. The result is added back to $X$ via a residual connection. This is followed by a feed-forward network (FFN), typically a two-layer MLP with a nonlinearity, and another residual connection. Layer normalization is applied around or before each block, depending on the variant. This repeated structure defines the backbone of LLMs and is where the bulk of activation memory resides—particularly at high sequence lengths, making it a natural target for low-rank approximation.

% \subsection{Transformer-based Large Language Models}
\subsection{Low-Rank Decomposition}
\label{subsec:lowrank_decomp}

\betterparagraph{Singular Value Decomposition(SVD)} 
Singular Value Decomposition decomposes a large 2D Matrix $T$ into the multiplication of three smaller matrices of $U$, $\Sigma$, and $V$. 
SVD is computed as:

\begin{equation}
T = U \times_{1} \Sigma \times_{2} V
\end{equation}
where \(U \in \mathbb{R}^{r_1 \times r_2}\), and \(U,~V\) belong to \(\mathbb{R}^{n_1 \times r_1},\mathbb{R}^{r_2 \times n_2}\) respectively.
$U$ is usually a tall, thin matrix, $V$ is a wide, short matrix, and $\Sigma$ is a square, diagonal matrix. 
Here, \(r_1,~r_2\) represent the decomposition rank of the matrix \(T\). Columns of matrix $U$ and rows of matrix $V$ are orthogonal and the singular values of matrix $\Sigma$ are sorted in Ascending order. The last values on $\Sigma's$ diagonal will be close to zero. The effect of this order on computation is that the first columns of the $U$ matrix and first rows of the $V$ matrix have larger impact on the original matrix $T$ reconstruction.
Thus, if we only pick the first $r_1$/$r_2$ columns/rows of matrix $U$/$V$, and the sub-matrix $[r_1, r_2]$ of matrix $\Sigma$ (low rank), we will be able to reconstruct the original matrix $T$ with reasonable approximation and low element-wise absolute error (MSE). The larger the $r_1$ and $r_2$, the lower the error and the higher the memory footprint and computation cost. 

%The \(i\) mode product, \(i = 1,~2,\), of the core matrix \(\Sigma\) and the factor matrices \(U^{1,~2,~3}\) is defined as: 
% 
%$$(U\times_1 \Sigma)(n_1,r_2) = \sum_{i_1=1}^{r_1} U(n_1,r_1)\Sigma(r_1,r_2)$$
% 
%$$(\Gamma \times_1 U^2)(r_1,n_2,r_3) = \sum_{i_2=1}^{r_2} \Gamma(r_1,i_2,r_3)U^1(i_2,n_2)$$
% 

\betterparagraph{Low-Rank SVD Approximation Error}
For a given set of decomposition ranks ($r_1$, $r_2$), the relative error between the original and the reconstructed matrix satisfies
\begin{equation}
\norm{T - (U\times_1\Sigma\times_2V)} \leq \epsilon \norm{T}
\end{equation}
where $\norm{T}$ is the \textit{norm} of $T$.
The goal of SVD is to minimize $\epsilon$, and it can be formulated as
\begin{equation}
\argmin_{\Sigma,U,V} \norm{T - (U\times_1\Sigma\times_2V)} 
\end{equation}
Generally, a higher decomposition rank results in a better approximation. While the lower bound of \(r_1,~r_2,~r_3\) is \(1\), the upper bound is usually taken as \(r_i = n_i,~i = 1,2,3\) for the optimal approximation. In our experiments, we prune the decomposition rank \(r_1 = r_2 = r_3 \in [1, \min(n_1,n_2,n_3)]\).

\subsection{SVD Computation Algorithms }
\label{subsec:svd_lanczos}
There are many algoritms to calculate SVD of a matrix/tensor. The most famous ones include QR Decomposition, Divide-and-Conquer, and Lanczos Algorithm. Although all of these algorithms eventually converge to the same decomposition and factor matrices, they differ in terms of convergence speed depending on the decomposition rank. For example, Divide-and-conquer is the fastest algorithm for relatively large ranks, but it is relatively slow for small ranks and requires a large memory footprint for large matrices. QR decomposition computes the precise factors and is faster than Divide-and-conquer for smaller ranks. Lanczos algorithm is the fastest for small ranks, since it iteratively constructs and refines the most important vectors/singular values. However, it is much slower for larger ranks and matrices.
Moar et al. characterize the design space of a low-rank decomposition application on a model's parameters. They demonstrate that small ranks are better choices in general since the decomposition effects on the model's quality do not differ significantly, while the memory footprint and computation reduction are significant for small ranks.
The goal of this work is to exert low-rank decomposition on activation matrices during real-time inference. Thus, we analyze the convergence speed of the SVD algorithms for small ranks to determine the best option in our use case.
~\autoref{fig:Decomposition_Algorithms_Convergence} compares the convergence speed of different SVD algorithms for different ranks. red dotted line demonstrates the optimal achievable decomposition using LAPACK routines. For small ranks, it is observed that Lanczos is considerably faster, suggesting its superiority in our use case. Accordingly, we target Lanczos for our activation decomposition. 

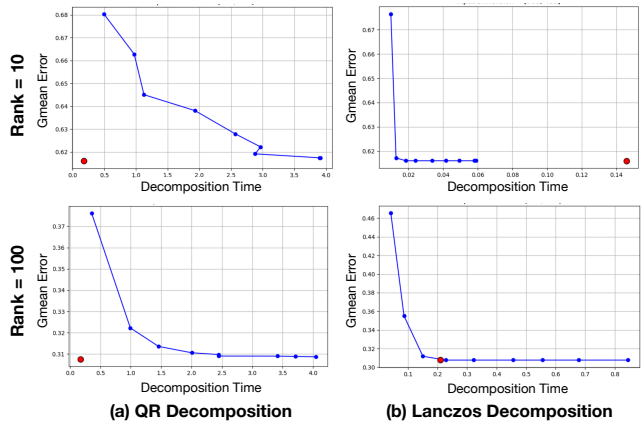
\insertFigure{Decomposition_Algorithms_Convergence}{Decomposition algorithms convergence speed on comparison a single A100 GPU. The input matrix size is [4096, 468].}

\betterparagraph{Lanczos Decomposition Algorithm and Runtime Analysis}
Lanczos is inherently an iterative algorithm, which constructs and refines factor matrices gradually. There are two versions of lanczos algorithm: Bidiagonalization and Tridiagonalization. We choose Lanczos Bidiagonalization algorithm since it doesn't need to perform $A^TA$ matrix multiplication and works directly on input matrix A. \autoref{alg:lanczos} is the pseudo code of Lanczos Bidiagonalization.

\noindent\rule{\linewidth}{1pt}

\begin{algorithm}
\caption{Lanczos Bidiagonalization}
\begin{algorithmic}[1]
\label{alg:lanczos}
\STATE Normalize $z_0$ and set $V[:,0] \gets z_0$
\STATE $u \gets A z_0$, $\alpha_0 \gets \|u\|$, $U[:,0] \gets u / \alpha_0$

\FOR{$j = 1$ to $k$}
    \STATE Orthogonalize $z \gets A^\top U[:,j-1]$ against $V$; 
           set $\beta_{j-1} \gets \|z\|$, $V[:,j] \gets z/\beta_{j-1}$
    \STATE Orthogonalize $u \gets A V[:,j]$ against $U$; 
           set $\alpha_j \gets \|u\|$, $U[:,j] \gets u/\alpha_j$
    \IF{$\alpha_j < \varepsilon$ or $\beta_{j-1} < \varepsilon$}
        \STATE \textbf{break}
    \ENDIF
\ENDFOR

\STATE Form bidiagonal $B$ from $\{\alpha,\beta\}$
\STATE Compute $(U_h, s, V_h) \gets \text{SVD}(B)$
\STATE Return $U U_h$, $s$, $V V_h^\top$
\label{code:lanczos}
\end{algorithmic}
\end{algorithm}
%\vspace{-7mm}
\noindent\rule{\linewidth}{1pt}
%\vspace{3mm}

%

\betterparagraph{Lanczos Bidiagonal Analysis}
~\autoref{fig:lanczos_bidiagonal_runtime_breakdown} provides details of running on a single A100 80GB GPU.
The runtime of all operations in the iterative algorithm has been shown. Amongst all, two operations of $U\ Reorthogonalization$ and $V\ Reorthogonalization$ take the majority of runtime, since they are in the most inner loop. These operations iteratively orthogonalize matrices multiple times to reduce the reconstruction error. Thus, they are not inherently parallelizable.

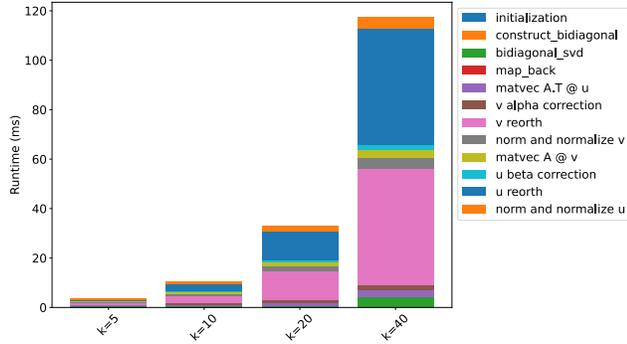
\insertFigure{lanczos_bidiagonal_runtime_breakdown}{Lanczos bidiagonal algorithm runtime breakdown on a single A100 GPU.}

\subsection{Low rank Decomposition on LLMs}
\label{subsec:low_rank_llms}
Low-rank decomposition has emerged as an effective technique for compressing and accelerating large language models (LLMs) by exploiting the observation that many weight matrices in transformers are highly redundant. Methods such as LoRA (Low-Rank Adaptation) by Hu et al. (2021) demonstrate that adapting only low-rank components of weight updates can significantly reduce the number of trainable parameters without sacrificing quality. Moar, C, et. al in \cite{IISWC_paper} fully characterize the design space of applying low-rank decomposition on language models. Specifically, they provides key insights about how to apply decomposition on model's weights to minimize the model's quality degradation.~\cite{tensorized} decompose embedding layers at the beginning of the model to reduce memory footprint. TIE~\cite{tie} also leverages Tensor Train decomposition for model compression in inference. The main advantages of low-rank decomposition include reduced memory footprint and faster inference on resource-constrained hardware, and enabling fine-tuning with limited compute. However, the approach also has trade-offs: aggressive rank reduction can lead to quality degradation, and the optimal rank choice may vary per layer and task.~\autoref{tab:related_works} summerizes the prior works.

\begin{table*}[ht]
\centering
\caption{Summary of prior works on low-rank decomposition.}
\label{tab:related_works}
\resizebox{\textwidth}{!}{%
\begin{tabular}{|l|c|c|c|}
\hline
\textbf{Work} & \textbf{Strategy} & \textbf{Accuracy Preservation Method} & \textbf{Goal} \\
\hline
LoRA~\cite{LoRA} & low-rank adaptation matrices for weight update & Fine-tuning & Parameter-efficient fine-tuning \\
\hline
Compressing Pre-trained LMs~\cite{compressing} & Apply SVD to weight matrices & Knowledge distillation & Memory reduction \\
\hline
Tensorized Embedding Layers~\cite{tensorized} & Decompose embedding layers into low-rank tensor factors & Jointly train factorized representation & Memory reduction \\
\hline
Holistic CNN Compression~\cite{holistic} & Apply Tucker/CP decomposition to convolutional kernels & Knowledge transfer & Latency/memory reduction \\
\hline
TIE~\cite{tie} & Replace dense layers with tensorized low-rank structures & Retraining after tensorization & Latency reduction \\
\hline
\hline
\accelerator \textit{(This work)} & Decompose Inputs and weights in decomposed-preserved format & Outlier-channel extraction & Latency and energy reduction  \\
\hline
\end{tabular}%
}
\end{table*}

\subsection{Motivation and Insights}
\label{subsec:motiv_insight}

Another orthogonal approach to apply low-rank decomposition on LLMs is to apply low-rank decomposition on inputs. However, In addition to the constraints mentioned,  the challenge is that trained model decomposition can be done offline and the new model will be deployed for use, while input decomposition at any stage (beginning of any layer) is a real-time process that should happen during inference and impose latency overhead if done naively and without algorithmic/hardware acceleration. Our goal is to propose a new computation graph and accelerator to achieve the minimal potential runtime.~\autoref{fig:Decomposition_Motivation} shows the comparison between the runtime of a single layer of Llama2-7b model during inference on 4 A100 80GB GPU against decomposition runtime on a single A100 GPU. We provided our proposed decomposer's runtime for comparison at a glance. We will discuss out methodology next.

\insertFigure{Decomposition_Motivation}{Comparison of Llama2-7b layer inference runtime for different input sequence lengths. Batch size is 64 and K(number of Lanczos iterations is 10).}
\section{Decomposition Methodology}
\label{method}

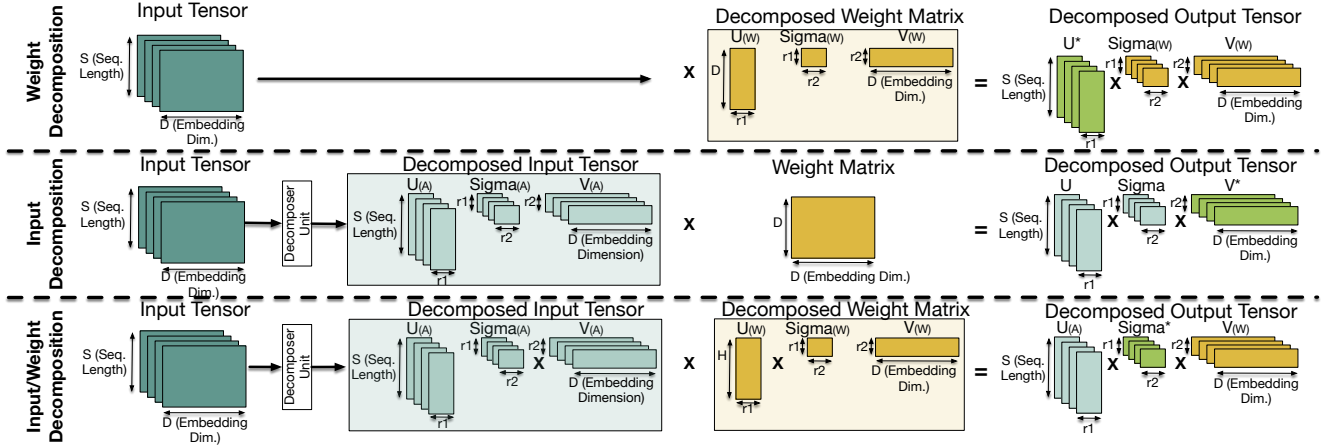
\insertWideFigure{Compute_Scheme}{Different decomposition strategies of a matmul layer in Large Language Models (LLMs) with output-decomposed computation.}

In this section, we explore low-rank decomposition on both activations and the model and the activations only. We elaborate on how it reduces model's computation runtime, and will formulate the computation and memory usage reductions of our method.

\subsection{Basic Decomposition Arithmetic}
\label{subsec:input_decomp_strategy}

\betterparagraph{Weight Decomposition}
Weight matrices of the pretrained model can be decomposed into low-rank factors and replace the original weight matrices. The literature shows that model decomposition is an enormous design space due to the large number of layers and ranks~\cite{IISWC_paper}, and it should be explored carefully to achieve the best compression with minimum model quality degradation. The new model can also be re-trained and slightly regain the quality. Applying low-rank decomposition on weights have been investigated in prior works, which we discussed in \autoref{subsec:low_rank_llms}.

\betterparagraph{Input Decomposition}
In this work, we explore input activation decomposition for language models. The dimensions of the input activation are batch\_size(B), Sequence length(S), and model's hidden dimension (H). 
The input to the model consists of $B$ prompts, each prompt is a 2D matrix with dimensions of $(S, H)$. 

To realize the input activation decomposition, we first divide the batch into separate prompts. Each 2D prompt is then passed to the SVD decomposition algorithm we deploy. After decomposition of all prompts, the factors and core matrices are concatenated to reconstruct the batch. \autoref{fig:Compute_Scheme} demonstrates the decomposition's input and output dimensions. Note that we apply the decomposition on each prompt separately, mainly because prompts typically do not have a meaningful relation. 
In computational perspective, the decomposition happen once at the beginning of the layer. The decomposed input is consumed by Query, Key, and Value matrix multiplication. The computation graph of a matmul with decomposed input is demonstrated in \autoref{fig:Compute_Scheme}b. instead of one large matmul, decomposed matmul includes three small matmuls, which reduces the number of FLOPs significantly for relatively small ranks. Although the order of matmul operations does not affect the output values, total number of computations and the average required memory footprint can vary significantly by changing the order of multiplication. Assuming thar $r_1, r_2 <<S, H$, The optimal computation order is:

\begin{equation}
U^{[S,r_1]} \times_3 \Sigma^{[r_1,r_2]} \times_2 V^{[r_2,H]} \times_1 W^{[H,H]}
\end{equation}

This will be done when a layer is chosen to be computed in a decomposed format, which requires the original output tensor. 

\betterparagraph{input+Weight Decomposition}
\label{subsec:input_weight_decomp_stratrgy}

We explore the combination of activation and weight decomposition in this work as well. The computation in this case will change to the following:

\begin{equation}
U_I^{[S,r_1]} \times_5 \Sigma_I^{[r_1,r_2]} \times_3 V_I^{[r_2,H]} \times_1 U_W^{[W,p_1]} \times_2 \Sigma_W^{[p_1,p_2]} \times_4 V_W^{[p_2,H]}
\label{eq:input_weight}
\end{equation}

where $p_1, p_2$ are the decomposition ranks of the weight matrix. As we discussed in \autoref{subsec:input_decomp_strategy}, the average required memory footprint varies considerably by changing the order of the multiplications. Again, assuming that $r_1, r_2, p_1, p_2 <<S, H$ and $p_1, p_2 < r_1, r_2$, performing matmuls as determined in \autoref{eq:input_weight} is the efficient order.

Applying decomposition to both weights and ifmaps has two major benefits. (1) The computation is significantly reduced even compared to the input-only decomposition. (2) the model itself will shrink, requiring less memory footprint and data transfer to compute units. However, it may amplify the negative effect on the model's quality. We will explore both approaches comprehensively in~\autoref{sec:evaluation}. 

There is a key challenge in input decomposition: reconstructed output computation. Assume that we aim to perform decomposed computation for an entire layer. We decompose ifmaps at the beginning of the layer for query, key, and value computation, but the output (ifmaps for attention score) will be in the original shape. Thus, we need to decompose it again before attention score computation. This process needs to be done after each matmul computation.
This has two crucial bottlenecks. First, the hardware resource requires the consideration of the same memory footprint as the original input, omitting the output tensor from the potential memory footprint reduction benefit. Second, the redundant decomposition for the upcoming computation is a significant burden on improving the latency. To resolve these two drawbacks, we propose $Output-Decomposed Computation$, which is explained in the following subsection.

\subsection{Decomposed-preserved Computation}

\betterparagraph{Input Decomposition method:}To address the challenges mentioned in~\autoref{subsec:input_weight_decomp_stratrgy}, we change the computation of the decomposed layer. Instead of conducting all three matrix computations, we only calculate the first matmul:

\begin{equation}
V^{*[r_2,H]} = V^{[r_2,H]} \times W^{[H,H]} 
\end{equation}

After the computation, a new $V*$ will be generated that can be associated with the input's $U$ and $\Sigma$ tensors to construct be the output of the block. This approach ensures that the output remains decomposed, and there will be no need to run the decomposition procedure before the next block. Similarly, if we decide to decompose consecutive layers, we use the same technique. However, the choice of decomposition layer should consider the quality of the model, as~\cite{IISWC_paper} shows that consecutive layer decomposition may negatively affect the accuracy of the model.

\betterparagraph{Input+Weight decomposition}
For weight and input decomposition, we only perform the first three matmuls.
\begin{equation}
\Sigma^{*[r_1, p_2]} = \Sigma_I^{[r_1,r_2]} \times_3 V_I^{[r_2,H]} \times_1 U_W^{[W,p_1]} \times_2 \Sigma_W^{[p_1,p_2]}
\end{equation}
Here, a new $\Sigma^*$ will be generated that can be associated with the input's $U$ and weight's $V$ tensors to construct the output. This can be directly used by the next matmul/layer.

Although decomposed-preserved computation reduces computation and memory footprint, keeping the outputs in decomposed format for many consecutive layers can affect model quality, since only one of the three factors keeps getting updated, while the other two remain intact. Thus, the decomposition error may accumulate and degrade the quality.
\section{Optimizing Model Quality: Outlier Extraction}
\betterparagraph{Outlier Definition}

\betterparagraph{Outlier Opportunity} Why outliers should be treated separately. Where are the outliers

To improve the model's quality, we extract outlier channels(columns) of the activations and separately decompose them in our computation graph. The decomposed outlier accompany the decomposed input in the computation path until we reconstruct the original activation map. \autoref{fig:Proposed_Computation} demonstrates an overview of our proposed decomposition scheme. We will discuss the details in the following.
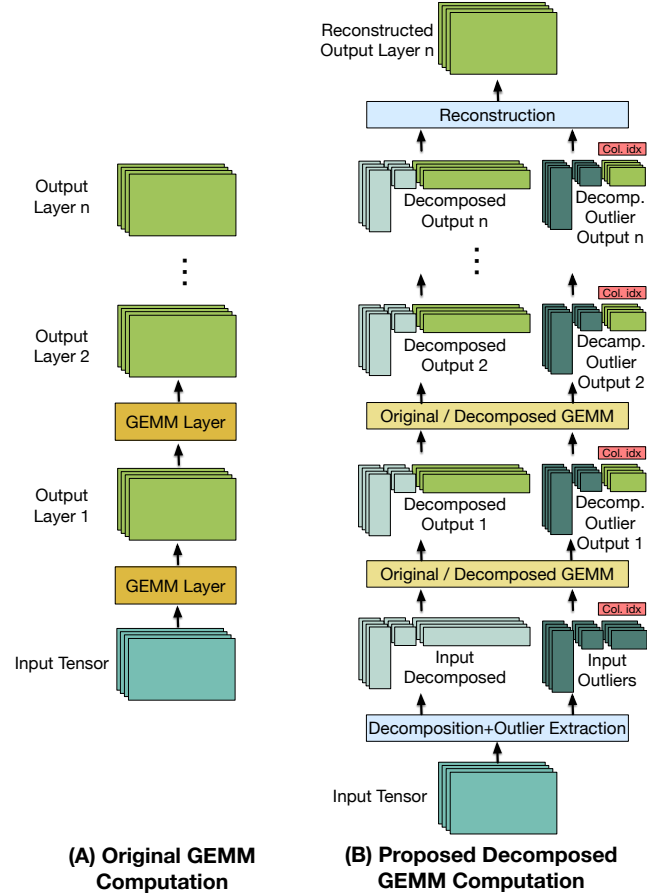
\insertFigure{Proposed_Computation}{Proposed computation scheme of GEMM/non-GEMM layers in the Model. (A) Original compute graph. (B) Decomposed and outlier-extracted compute graph.}
Research shows that classical low-rank decomposition techniques such as SVD perform best when the input data is distributed relatively uniformly and free of extreme values\cite{robust_svd, robust_pca}. Because these methods minimize squared reconstruction error, they are highly sensitive to outliers–even a small fraction of large errors can disproportionately alter the recovered subspace directions\cite{robust_pca}. 
In this paper's context, model inputs inherently include small number of outliers in the activation map, which makes the data distribution not ideal for low-rank decomposition, especially if ranks are very small (close to 1). 
To address this, we aim to separate out the outliers from the activation map before applying low-rank decomposition. However, element-wise outlier extraction from a large activation map and storing them using metadata is not a cheap computation in terms of latency and energy. To determine the methodology and granularity of outlier extraction, we need to analyze the activation maps of different layers in detail.
\autoref{fig:activation_map_layers} depicts the activation map values of a sample prompt for four different layers. The observation is that outliers are not randomly distributed. They mainly reside in specific channels (corresponding to the hidden dimension "$H$") and a few specific tokens (corresponding to the hidden dimension "$S$"). To minimize the outlier extraction overhead, we apply channel-wise outlier extraction.
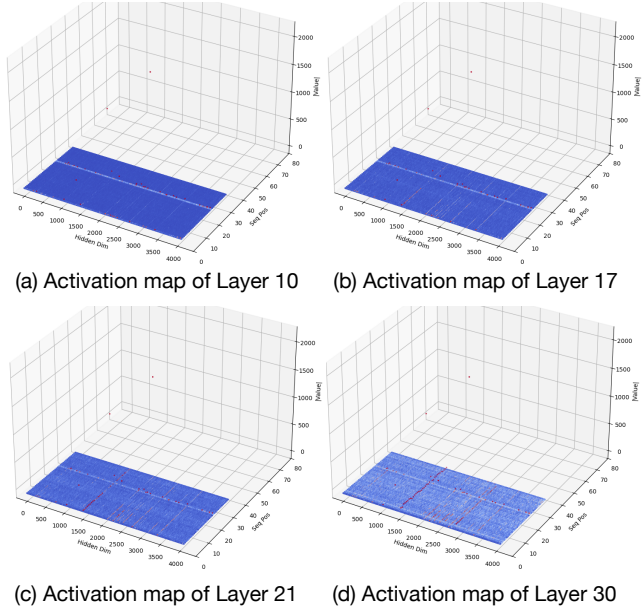
\insertFigure{activation_map_layers}{Activation map of four layers in Llama-2-7b. Red dots demonstrate higher absolute values, and blue dots indicate small absolute values.}
Specifically, we detect the channel to be considered as an outlier by counting the number of outlier elements. The algorithm specifies a threshold $T$ to determine if a value is an outlier or not. This threshold is calculated based on an offline analysis of the input feature map of the model's intermediate layers. 
Our observation shows that a statically-determined threshold by various workloads and benchmarks can capture a reasonably small number of channels for all workloads. However, the feature map values vary for the inputs at each layer, and outliers cannot be captured using a unified threshold. Thus, a table including the outlier thresholds for each layer in the model is created offline using statistical analysis. When a layer is chosen for input decomposition, the outlier extraction algorithm uses the threshold corresponding to that layer. The percentage of outlier extraction for different layers and workloads vary from 5.05\% to 2.12\% and the average is 3.02\%
%
%to decompose the origina low-rank component plus a sparse outlier component, explicitly modeling and separating high-magnitude anomalies. Central to their guarantees is an incoherence assumption, which ensures the low-rank signal is spread across entries rather than concentrated in a few corrupted columns or rows (ar5iv, arXiv, PMC). More recent advances—such as spherically normalized SVD—achieve significant robustness improvements (higher breakdown points) with runtime comparable to classical methods (ar5iv, arXiv). In other words, accurate low-rank decomposition hinges on having a data distribution that is not dominated by outliers, or else robustness-enhanced methods must be employed.
%Table\autoref{decomp_only_results} demonstrate the model's performance for different decomposition configurations

\subsection{Computation and memory footprint reduction}

\betterparagraph{Computation Analysis} We formalize the computation reduction for input-only decomposition and input-weight decomposition. For input-only decomposition, the computation reduction is calculated as:

%$$\# ~MACs~before~decomposition = B \times S \times H \times W$$
%$$\#~MACs~after~decomposition = B \times (r_2 \times H \times W )$$
\begin{equation}
Compute\ Reduction\ Ratio = \frac{B \times S \times D \times W}{B \times r_2 \times D \times W} = \frac{S}{r_2}
\end{equation}

For input-weight decomposition, the computation reduction is calculated as:

%$$\#~MACs~after~decomp. = B \times (r_2 \times H \times p_1 + r_2 \times p_1 \times p_2 + r_1 \times r_2 \times p_2)$$

$$\ Compute\ Reduction\ Ratio = $$
\begin{equation}
\frac{S \times D \times W}{r_2 \times D \times p_1 + r_2 \times p_1 \times p_2 + r_1 \times r_2 \times p_2}
\end{equation}

\betterparagraph{Memory Footprint} We break down memory footprint into two parts. First is the required memory for input activation storage, and second, the required memory to store model parameters.

For input activations, we assume that \(r_1, r_2 < \min(H,~W)\). The required memory to store input activations, and the memory reduction ratio can be computed as:
%$$\# ~memory~before~decomposition = B \times S \times H$$
%$$\#~Memory~after~decomposition = B \times (S \times p_1 + p_1 \times p_2 + p_2 \times H)$$
\begin{equation}
Compression~Ratio = \frac{S \times D}{S \times p_1 + p_1 \times p_2 + p_2 \times D}
\end{equation}

To formulate the required memory for parameters, we assume that \(r_1, r_2, p_1, p_2 < \min(H,~W)\). The number of parameters (relative to memory footprint) will reduce if:

\begin{equation}
(p_1, p_2 < (\frac{\sqrt{(D+W)^2 + 4\times D\times W} - (D + W)}{2}))
\end{equation}

More specifically, the total number of parameters is reduced due to decomposition, and the compression ratio can be computed as:
%$$\# ~params~before~decomposition = H \times W$$
%$$\#~params~after~decomposition = H \times p_1 + p_1 \times p_2 + p_2 \times W$$

\begin{equation}
Compression~Ratio = \frac{D \times W}{D \times p_1 + p_1 \times p_2 + p_2 \times W}
\end{equation}

\section{Decomposer Accelerator}
\label{sec:accelerator}

In this section, we first characterize and profile the computational overhead of Lanczos algorithm used for decomposition, then we propose our architecture and computation scheme that meets real-time the requirement of real-time input activation decomposition.

\subsection{\accelerator Architecture}
\label{subsec:Architecture}

\accelerator is structured with multiple clusters organized around distributed memory banks, forming a scalable and highly parallel accelerator design. \autoref{fig:Accelerator} provides an overview of the proposed architecture. Specifically, \accelerator consists of 256 clusters arranged in a $16 \times 16$ two-dimensional array. Each column of clusters is paired with a dedicated memory bank, responsible for storing and streaming a partition of the vector data to the compute units. This partitioning is particularly effective for iterative vector operations, since it minimizes global memory accesses and improves locality of reference.

The architecture is designed to be flexible and composable, such that \accelerator can be deployed alongside conventional GEMM accelerators, including commercial GPUs such as NVIDIA A100 or H100~\cite{nvidia_h100, a100}, as well as future specialized accelerators. Leveraging the iterative computation expansion methodology discussed in \autoref{subsec:compute_expansion}, the scale of \accelerator has been carefully selected: 256 clusters are sufficient to decompose and process any input size across large-scale models, while still ensuring faster execution than the baseline GEMM runtime on a 4-rank A100 GPU system.

From a hardware cost perspective, a single \accelerator core occupies nearly $7\times$ less area compared to a core with equivalent compute capability in an A100 GPU. This emphasizes the efficiency of \accelerator as a complementary accelerator for end-to-end runtime improvement, enabling both performance gains and hardware savings. In the following subsection, we describe the internal cluster organization that makes this efficiency possible.

\subsection{Cluster}
\accelerator’s cluster is the fundamental compute building block, consisting of 64 FP16 multipliers arranged in an $8 \times 8$ two-dimensional array. Each cluster is equipped with a shared buffer that stores the local data partition assigned to that cluster. This distributed buffering strategy provides a key advantage over a unified memory system: it offers higher effective memory bandwidth to the compute units by reducing contention and bringing data closer to computation.

To accelerate iterative vector operations, each cluster integrates a network of reduction and scatter units. Specifically, every multiplier is connected to two independent reduction paths: one horizontal (row-wise) and one vertical (column-wise). These reduction units are implemented as binary-tree structures, enabling logarithmic-depth reduction operations and therefore minimizing latency during collective computations. This design is particularly well-suited for repeated decomposed operations where reductions dominate the workload.

Together, the $8 \times 8$ multiplier array, the shared buffer, and the dual-path reduce/scatter network form a highly efficient compute cluster. \autoref{subsec:compute_expansion} further elaborates on how these architectural choices map naturally to iterative decomposed workloads, while \autoref{fig:Accelerator} illustrates the detailed structure of a single cluster.
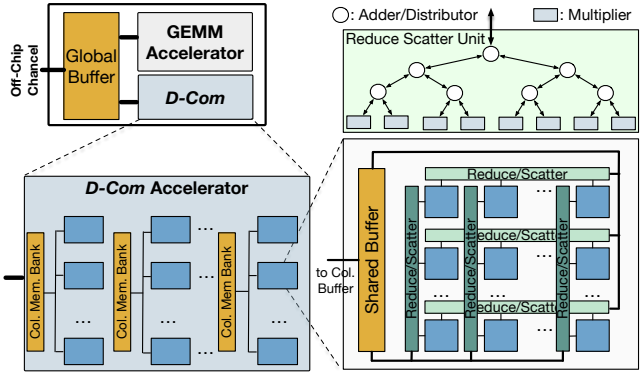
\insertFigure{Accelerator}{An overview of the \accelerator architecture, showing the global 16$\times$16 cluster arrangement and the internal organization of each cluster unit.}

\subsection{Computation Expansion and Mapping}
\label{subsec:compute_expansion}

\autoref{fig:Lanczos_Compute_Expansion} (a) depicts the straightforward computation graph and the hardware mapping of two operations mentioned in~\autoref{subsec:svd_lanczos}. Depending on the hardware, the process can happen within multiple SMs in GPU, or within Vector Processor Unit(VPU) in TPUs. The computation involves data read from memory, parallel multiplication, vector reduction (orange arrows), broadcast, parallel multiplication and subtraction, and memory write-back. The main latency bottleneck is memory read/write and vector reduction for large vectors.

To improve the latency, we propose $Computation\ Expansion$. The intuition behind $Computation\ Expansion$ is that iterative vector operations are memory-bound and most compute units will be idle during these processes. If we employ more compute units and provide sufficient bandwidth for all units, we can accelerate the iterative algorithms. These features are realized in \accelerator.
More specifically, we can omit or shorten the vector reduction in~\autoref{fig:Lanczos_Compute_Expansion}a and broadcast the partial products to the next element-wise multiplication. The next element-wise multiplication needs to be duplicated if we want to parallelize their computation.

\autoref{fig:Lanczos_Compute_Expansion}b depicts the fully expanded computation graph. Although fully expanding the computation changes the nature of the algorithm from memory-bound to compute-bound, improving the latency is not guaranteed since computation overhead may exceed the memory transfer improvement. Moreover, we eventually need to aggregate all partial results of the correction vector at the end (blue arrows), which vanishes the vector reduction benefit of computation expansion and wastes energy. 

Instead, we can partially expand the computation. \autoref{fig:Lanczos_Compute_Expansion}c illustrates an example of partially expanded computation. This divides the reduction into two parts. As seen in computation mapping of~\autoref{fig:Lanczos_Compute_Expansion}c, both reductions are localized among 4 cores (2-in-2 squares). Another crucial improvement is that both $V$ and $z$ vectors can be distributed among squares. Although we still need global broadcast, it can happen by one consecutive write and read on a small global memory for broadcast purposes.

Finding the optimal expansion factor depends on the accelerator's scale. Depending on the desired speedup, hardware scale and expansion factor can be optimized. in~\autoref{subsec:latency_comparison}, we measure various expansion factors and find the optimal one for our target \accelerator scale.

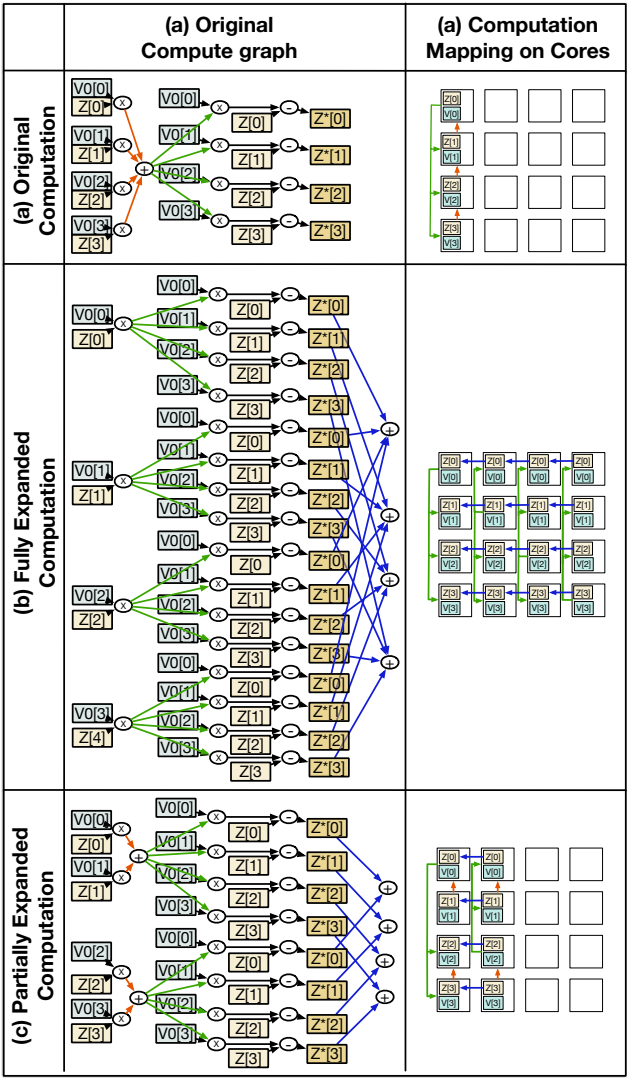
\insertFigure{Lanczos_Compute_Expansion}{Re-orthogonalization of V computation graph as the latency bottleneck of Lanczos bidiagonalization algorithm.}

\section{Evaluation}
\label{sec:evaluation}

\subsection{Evaluation Methodology}

We use pretrained Llama-2-7b from huggingface repository~\cite{llama-2-7b} as our experimental model. The model's runtime measurements are based on our 4 A100 80GB GPUs. Our evaluation datasets are arc\_easy and wikitext-2. Accuracy is used for arc\_easy and perplexity is used for wikitext-2 as the metric. We develope RTL implementation of \accelerator in System Verilog and synthesize it using 15 nm technology \cite{martins2015open} for area and power analysis. For latency comparison, we develope a performance model for both \accelerator and A100 GPU and validated the results with the actual A100 runtime.

We implement the RTL design of \accelerator in System Verilog. We synthesize the implementation with Synopsys Design Compiler using a 15 nm technology library to evaluate the area and power costs. We also model the quality of \accelerator for iterative algorithms, specifically Lanczos Bidiagonalization, and compare it against A100 80GB GPU runtime with an equal amount of compute and memory resources.

\subsection{Decomposition Configuration Exploration}

We evaluate the impact of input activation decomposition and input+model decomposition on model quality. We run 4 various layer choices for decomposition. We inspired from prior work~\cite{IISWC_paper} regarding the choice of layers for decomposition. The model quality is maintained better if the decomposed layers are not adjacent. We also experiment with 3 different ranks (1, 10, and 20) for all decompositions. Also, we keep the inputs decomposed for all matmuls within a layer. We should note that the decomposition choice is an extremely large design space that can be explored further in future research.

\begin{table*}[ht]
\label{tab:result_input_decomp}
\centering
\caption{Input decomposition results. Accuracy and perplexity are based on arc\_easy and wikitext2, respectively. All other results are based on running arc\_easy dataset. Total runtime is reported with D-com deployment. The found configuration with the best speedup-quality trade off is highlighted.}
\begin{adjustbox}{width=\textwidth}
\begin{tabular}{|l|c|c|c|c|c|c|c|c|c|}
\hline
\textbf{Decomposed} & \textbf{Decomp.} & \textbf{Outlier} & \textbf{Accuracy\% /} & \textbf{Model} & \textbf{Decomp.} & \textbf{Decomp.} & \textbf{Memory} & \textbf{Total Runtime} \\
\textbf{Layers} & \textbf{Rank} & \textbf{Extraction \%} & \textbf{Perplexity} & \textbf{Runtime} & \textbf{GPU Time} & \textbf{Accel. Time} & \textbf{Reduction \%} & \textbf{Reduction} \\
\hline
Original & - & - & 73.8 / 11.57 & 1x(122 s) & - & - & - & - \\
\hline
[10, 15, 20, 25] & 1 & 4.0\% & 68.98 / 17.72 & 0.90x & 14.2 s & 1.8 s & 9.5\% & 10\% \\
\hline
[10, 15, 20, 25] & 10 & 3.0\% & 70.8 / 15.93 & 0.91x & 25.1 s & 3.2 s & 8.8\% & 9\% \\
\hline
[10, 15, 20, 25] & 20 & 2.9\% & 72.7 / 13.81 & 0.92x & 42.6 s & 5.3 s& 7.4\% & 8\% \\
\hline
[6, 10, 14, 18, 22, 26] & 1 & 4.1\% & 64.1/25.76 & 0.86x & 21.3 s & 2.6 s & 13.1\% & 14\% \\
\hline
[6, 10, 14, 18, 22, 26] & 10 & 3.1\% & 70.6 / 16.41 & 0.87x & 37.6 s & 4.7 s& 12.2\% & 13.0\% \\
\hline
[6, 10, 14, 18, 22, 26] & 20 & 2.9\% & 72.7 / 13.66 & 0.88x & 64.0 s & 8.1 s& 11.1\% & 12.0\% \\
\hline

[7, 10, 13, 16, 19, 22, 25, 28] & 1 & 4.0\% & 62.4 / 48.58 & 0.82x & 27.8 s & 3.5 s& 17.1\% & 18.1\% \\
\hline
[7, 10, 13, 16, 19, 22, 25, 28] & 10 & 3.1\% & 68.0 / 19.28 & 0.84x & 49.2 s & 6.2 s& 15.8\% & 16.4\% \\
\hline
[7, 10, 13, 16, 19, 22, 25, 28] & 20 & 2.9\% & 71.5 / 16.14 & 0.85x & 85.1 s & 10.6 s& 14.3\% & 15.7\% \\
\hline
[9, 10, 13, 14, 17, 18, 21, 22, 26, 27] & 1 & 4.1\% & 57.57 / 47.20 & 0.74x & 35.5 s & 4.43 s& 24\% & 26\% \\
\hline
[9, 10, 13, 14, 17, 18, 21, 22, 26, 27] & 10 & 3.2\% & 63.00 / 28.76 & 0.76x & 62.4 s& 7.8 s & 22.9\% & 24\% \\
\hline
\hline
\textbf{[9, 10, 13, 14, 17, 18, 21, 22, 26, 27]} & \textbf{20} & \textbf{2.9\%} & \textbf{70.15 / 17.03} & \textbf{0.78x} & \textbf{103.7 s} & \textbf{13.0 s} & \textbf{21.7\%} & \textbf{22\%} \\
\hline
\hline
All Layers (Most aggressive) & 1 & 6.5\% &  26.58 /168218 & 0.35x & 113.0 s & 14.1 s& 71.4\% & 65\% \\
\hline

\end{tabular}
\end{adjustbox}
\end{table*}

\begin{table*}[ht]
\label{tab:result_input_weight_decomp}
\centering
\caption{Input + Weight decomposition results. Accuracy and perplexity are based on arc\_easy and wikitext2, respectively. All other results are based on running arc\_easy dataset. Total runtime is reported based on \accelerator deployment.}
\begin{adjustbox}{width=\textwidth}
\begin{tabular}{|l|c|c|c|c|c|c|c|c|c|}
\hline
\textbf{Decomposed} & \textbf{Decomp.} & \textbf{Outlier} & \textbf{Accuracy /} & \textbf{Model} & \textbf{Decomp.} & \textbf{Decomp.} & \textbf{Memory Reduction} & \textbf{Total Runtime} \\
\textbf{Layers} & \textbf{Rank} & \textbf{Extraction \%} & \textbf{Perplexity} & \textbf{Runtime} & \textbf{GPU Time} & \textbf{Accel. Time} & \textbf{(input/weight)} & \textbf{Reduction \%} \\
\hline
Original & - & - & 73.8 / 11.57 & 1x(122 s) & - & - & - & - \\
\hline
[10, 15, 20, 25] & 1 & 4.1\% & 67.04 / 16.57 & 0.88x & 14.2 s& 1.8 s& 9.5\% / 12.0\% & 12\% \\
\hline
[10, 15, 20, 25] & 10 & 3.5\% & 66.7 / 15.88 & 0.89x & 25.1 s & 3.2 s &  9.5\% / 11.9\% & 11\% \\
\hline
[10, 15, 20, 25] & 20 & 3.3\% & 66.9 / 15.61 & 0.89x & 42.6 s & 5.3 s &  9.5\% / 11.9\% & 10\% \\
\hline
[6, 10, 14, 18, 22, 26] & 1 & 4.2\% & 60.2 / 23.50 & 0.83x & 21.3 s & 2.6 s &  13.1\% / 18.0\% & 16.7\% \\
\hline
[6, 10, 14, 18, 22, 26] & 10 & 3.8\% & 59.21 / 27.21 & 0.84x & 37.6 s & 4.7 s &  12.2\% / 17.9\% & 15.3\% \\
\hline
[6, 10, 14, 18, 22, 26] & 20 & 3.5\% & 60.58 / 25.04 & 0.84x & 64.0 s & 8.1 s &  11.1\% / 17.8\% & 13.8\% \\
\hline
[7, 10, 13, 16, 19, 22, 25, 28] & 1 & 4.1\% & 54.75 / 51.83 & 0.80x & 27.8 s & 3.5 s &  17.1\% / 24.0\% & 20\% \\
\hline
[7, 10, 13, 16, 19, 22, 25, 28] & 10 & 3.7\% & 52.86 / 58.09 & 0.82x & 49.2 s & 6.2 s &  15.8\% / 23.8\% & 18\% \\
\hline
[7, 10, 13, 16, 19, 22, 25, 28] & 20 & 3.5\% & 52.56 / 57.11 & 0.84x & 85.1 s & 10.6 s &  14.3\% / 23.7\% & 16\% \\
\hline
[9, 10, 13, 14, 17, 18, 21, 22, 26, 27] & 1 & 4.0\% & 48.98 / 65.33 & 0.71x & 35.5 s & 4.4 s &  24\% / 30.0\% & 29\% \\
\hline
[9, 10, 13, 14, 17, 18, 21, 22, 26, 27] & 10 & 3.7\% & 46.54 / 72.71 & 0.73x & 62.4 s & 7.8 s &  22.9\% / 29.5\% & 27\% \\
\hline
[9, 10, 13, 14, 17, 18, 21, 22, 26, 27] & 20 & 3.5\% & 45.95 / 74.92 & 0.75x & 103.7 s & 13.0 s &  21.7\% / 29.3\% & 25\% \\
\hline
All Layers (Most aggressive) & 1 & 4\% & 25.92 / $7 \times 10^6$ & 1.20x & 113.0 s & 14.1 s &  71.4\% / 96\% & 74\% \\
\hline
\end{tabular}
\end{adjustbox}
\end{table*}

Our experimental results in \autoref{tab:result_input_decomp} indicate a trade-off between model quality and computational efficiency as the number of decomposed layers increases. Specifically, decomposing more layers leads to notable improvements in both runtime and memory footprint, particularly when using \accelerator for decomposition. The total runtime benefits from the significant latency reduction of selected layers for input decomposition. Memory usage is reduced by $15.6\%$ on average. However, this efficiency gain comes at the cost of a modest decline in model accuracy and an increase in perplexity. This degradation becomes pronounced as more layers are decomposed due to the compounded approximation error introduced by low-rank representations. Outlier extraction mitigates this effect significantly to isolate and preserve the most expressive components of the activations before decomposition. It improves the fidelity of the approximated tensors. Overall, the results demonstrate the potential of low-rank decomposition with targeted outlier handling to balance latency, memory, and accuracy in large language models.

\autoref{tab:result_input_weight_decomp} demonstrates Input+Model decomposition. If we compare the corresponding numbers with~\autoref{tab:result_input_decomp}, we see a better memory footprint and reduced latency. However, the model's quality is also affected due to the error multiplication of the decomposed input and weight values. Athough the number of computations significantly reduce in input+weight decomposition, the runtime is not meaningfully better than input-only decomposition. The reason is multiple small matrix multiplication. Similar to vector operations, small matrix multiplications are also memory-bound and cannot benefit from reducing computation after a certain point.

\subsection{Outlier Extraction Effect on Model Quality}
We study the effectiveness of outlier extraction on input-decomposed method as the superior method in terms of accuracy. \autoref{fig:Outlier_Effect} illustrates the impact of outlier extraction effect on input-decomposed method for different ranks. Extracting $3\%$ of outliers on average can considerably improve model quality. However, going beyond $5\%$ cannot significantly elevate the performance while imposing computation overhead and diminish the decomposition latency and memory benefits.~\autoref{fig:Outlier_Effect} visualizes the outlier extraction impact for different ranks. We experiment outlier percentage analysis on 4-layer decomposition configuration.

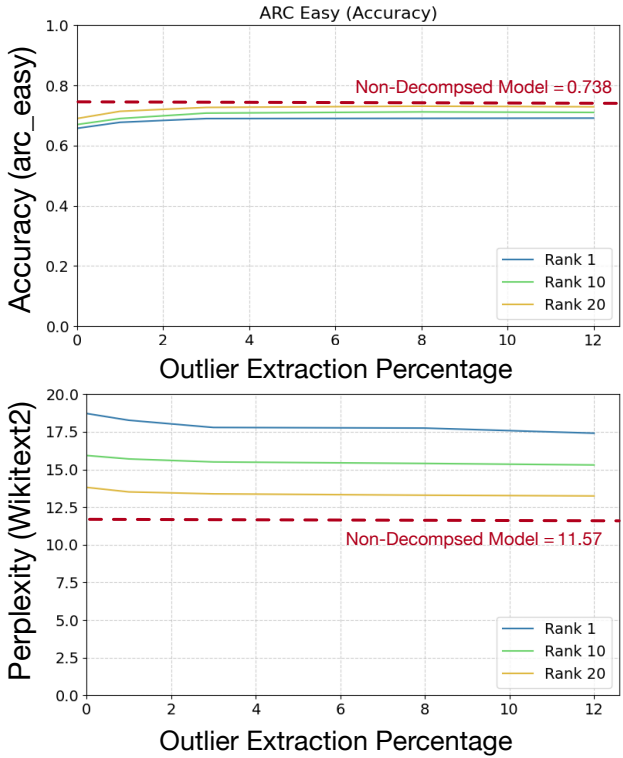
\insertFigure{Outlier_Effect}{Outlier extraction effect for different ranks. The number of decomposed layers are 4.}

\subsection{\accelerator Simulation: Model quality and latency}
\label{subsec:latency_comparison}

Based on our decomposition config exploration, we found several promising configs with considerable runtime improvement potential and tolerable accuracy loss. We choose the highlighted configuration in~\autoref{tab:result_input_decomp} as our best configuration and use it for our latency evaluation and comparison. \autoref{fig:Runtime_Comparison}a compares the original layer runtime, decomposed layer runtime on A100, and decomposed model runtime on \accelerator. When the input decomposition is done naively on the same hardware, not only the decomposition benefit vanishes, but also the overhead results in $2.3\times$ more latency. Deploying \accelerator, the decomposition is about $8\times$ faster and is realized on the dedicated accelerator. Since the speedup is sufficient enough to run in parallel with both original and decomposed layers, the latency improvement is $3.8\times$ less than original layer, and $8.74\times$ better than decomposed layer on A100. \autoref{fig:Runtime_Comparison}b demonstrated the end-to-end model latency comparison including decomposed and non-decomposed layers. In terms of model quality, the original accuracy and perplexity on arc\_easy and wikitext is 73.8\% and 11.6, respectively, and for the best configuration, the accuracy and perplexity of the decomposed model is 70.2\% and 17.0, respectively. 

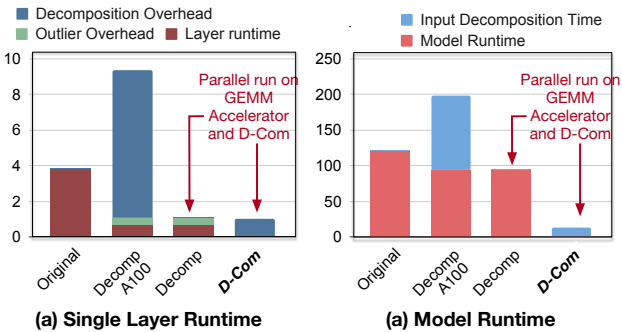
\insertFigure{Runtime_Comparison}{Runtime Comparison of original model, decomposed model on A100, and decomposed model on \accelerator. (a) is a single layer runtime, and (b) is the entire model runtime for our best decomposition configuration (see~\autoref{tab:result_input_decomp}).}

For any \accelerator scale, there is an optimized expansion factor that resulrs in the optimal latency. For our chosen scale, the optimized expansion factor is 8.\autoref{fig:Latency_Comparison} provides the results for various expansion factors. For $f=8$, the computation and memory transfer reach to a balanced point, where the accelerator exploits the maximum memory bandwidth and compute resources. For $f$ smaller than 8, the iterative algorithm is still memory-bound, and for $f$ larger than 8, the algorithm becomes compute-bound, meaning that the accelerator does not have sufficient cores to expand the computations with that factor. $f$ can also be determined based on the model designer's acceleration requirement to prevent unnecessary speedup and save more energy.

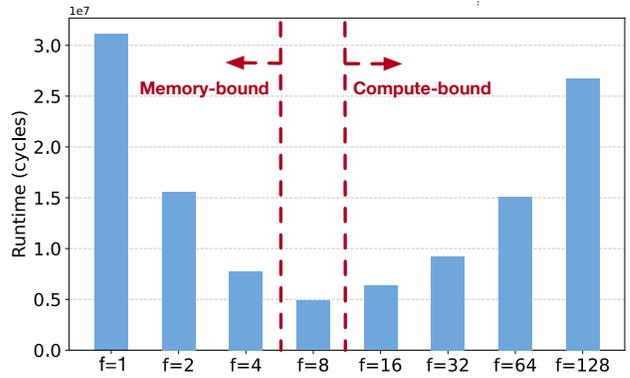
\insertFigure{Latency_Comparison}{Decomposition latency comparison of \accelerator for different expansion factors (f). The Batch size is 64, Sequence length is 4096, embedding dim. is 4096, and decomposition rank is 10. D-Com scale is as described in~\autoref{subsec:Architecture}}

\subsection{Area and Power}
Our proposed scale for \accelerator which has $16 \times 16$ clusters and $8 \times 8$ MACs within each cluster is capable of meeting the realtime parallel decomposition requirement. This scale is roughly $7 \times$ smaller than an accelerator with the same compute capability as a single A100 GPU. This clarifies the area and power efficiency of our methodology for LLM speedup. \autoref{fig:Area_Power} demonstrates the area and power comparison of \accelerator against a systolic array with the same compute capability and memory. Our area is $3\%$ higher than a typical systolic array. However, our power consumption is $59\%$ less than systolic array due to less global on-chip communications and distributed memory.

\insertFigure{Area_Power}{Area and power comparison of \accelerator against TensorCore}
\section{Related Works}
\label{sec:related_works}

Hu et al.\cite{LoRA} proposes LoRA that enables fine-tuning transformers with updating only a fraction of low-rank parameters. Inspired by similar principles, AdaLoRA\cite{adalora} and LoTR\cite{lotr} introduce adaptive rank allocation to better capture task-specific importance during fine-tuning.
Moar, C, et. al in~\cite{IISWC_paper}
characterize the design space of applying low-rank de-
composition on LLM's weights to achieve speedups while minimize the model’s quality degradation
.TIE framework\cite{tie} presents an inference-friendly method for accelerating deep neural networks by leveraging Tensor Train decomposition for model compression. In a complementary direction, Saha et al.\cite{low_prec_low_rank} propose approximating model weights through a hybrid representation, where a low-precision matrix is combined with a low-rank high-precision matrix, effectively balancing efficiency with accuracy.
~\cite{tensorized} proposes input embedding layer decomposition using Tensor Train decomposition. However, it does not effectively reduce memory footprint or latency since the rest of the layers are computed in the original shape.
Kopiczko et al.~\cite{Kopiczko} investigate decomposed fine-tuning strategies that jointly optimize rank and scaling factors to balance accuracy and efficiency.
The authors in~\cite{stable} apply Canonical Polyadic (CP) low rank decomposition on CNNs.
They investigate the effectiveness of Tucker and CP decomposition combination for convolutional layers in CNNs.

\section{Conclusion}
\label{sec:conclusion}

Motivated by the heavy compute- and memory-overheads of LLMs, many model compression techniques have been explored.
Among them, activation decomposition has not been actively explored since the runtime overhead of decomposition often exceeds the benefits in off-the-shelf hardware options.
In this work, we show that activation decomposition can actually be a good option with a proper choice of decomposition algorithm, hardware support, and co-design of algorithm and hardware.
We also show the efficacy of compute expansion methodology, which mitigates the memory boundness with carefully mapped replicated computations.
We believe such an approach can be a major breakthrough for memory-bound operations commonly found in recent LLM workloads, which we expect follow-up studies.

%%%%%%% -- PAPER CONTENT ENDS -- %%%%%%%%

%%%%%%%%% -- BIB STYLE AND FILE -- %%%%%%%%
%\bibliographystyle{IEEEtranS}
%\bibliography{refs}
\bibliographystyle{ACM-Reference-Format}
\bibliography{main}


\begin{thebibliography}{24}

%%% the Web address.

\ifx \showCODEN    \undefined \def \showCODEN     #1{\unskip}     \fi
\ifx \showDOI      \undefined \def \showDOI       #1{#1}\fi
\ifx \showISBNx    \undefined \def \showISBNx     #1{\unskip}     \fi
\ifx \showISBNxiii \undefined \def \showISBNxiii  #1{\unskip}     \fi
\ifx \showISSN     \undefined \def \showISSN      #1{\unskip}     \fi
\ifx \showLCCN     \undefined \def \showLCCN      #1{\unskip}     \fi
\ifx \shownote     \undefined \def \shownote      #1{#1}          \fi
\ifx \showarticletitle \undefined \def \showarticletitle #1{#1}   \fi
\ifx \showURL      \undefined \def \showURL       {\relax}        \fi
% invisible to TeX
\providecommand\bibfield[2]{#2}
\providecommand\bibinfo[2]{#2}
\providecommand\natexlab[1]{#1}
\providecommand\showeprint[2][]{arXiv:#2}

\bibitem[Anthropic(2023)]%
        {claude2}
\bibfield{author}{\bibinfo{person}{Anthropic}.} \bibinfo{year}{2023}\natexlab{}.
\newblock \bibinfo{title}{Model Card and Evaluations for Claude Models}.
\newblock
\newblock
\urldef\tempurl%
\url{https://www.anthropic.com/news/claude-2}
\showURL{%
\tempurl}


\bibitem[Bershatsky et~al\mbox{.}(2024)]%
        {lotr}
\bibfield{author}{\bibinfo{person}{Daniel Bershatsky}, \bibinfo{person}{Daria Cherniuk}, \bibinfo{person}{Talgat Daulbaev}, \bibinfo{person}{Aleksandr Mikhalev}, {and} \bibinfo{person}{Ivan Oseledets}.} \bibinfo{year}{2024}\natexlab{}.
\newblock \showarticletitle{{LoTR}: Low tensor rank weight adaptation}.
\newblock \bibinfo{journal}{\emph{arXiv [cs.CL]}} (\bibinfo{date}{Feb.} \bibinfo{year}{2024}).
\newblock


\bibitem[Brown et~al\mbox{.}(2020)]%
        {few_shots}
\bibfield{author}{\bibinfo{person}{Tom Brown}, \bibinfo{person}{Benjamin Mann}, \bibinfo{person}{Nick Ryder}, \bibinfo{person}{Melanie Subbiah}, \bibinfo{person}{Jared~D Kaplan}, \bibinfo{person}{Prafulla Dhariwal}, \bibinfo{person}{Arvind Neelakantan}, \bibinfo{person}{Pranav Shyam}, \bibinfo{person}{Girish Sastry}, \bibinfo{person}{Amanda Askell}, \bibinfo{person}{Sandhini Agarwal}, \bibinfo{person}{Ariel Herbert-Voss}, \bibinfo{person}{Gretchen Krueger}, \bibinfo{person}{Tom Henighan}, \bibinfo{person}{Rewon Child}, \bibinfo{person}{Aditya Ramesh}, \bibinfo{person}{Daniel Ziegler}, \bibinfo{person}{Jeffrey Wu}, \bibinfo{person}{Clemens Winter}, \bibinfo{person}{Chris Hesse}, \bibinfo{person}{Mark Chen}, \bibinfo{person}{Eric Sigler}, \bibinfo{person}{Mateusz Litwin}, \bibinfo{person}{Scott Gray}, \bibinfo{person}{Benjamin Chess}, \bibinfo{person}{Jack Clark}, \bibinfo{person}{Christopher Berner}, \bibinfo{person}{Sam McCandlish}, \bibinfo{person}{Alec Radford}, \bibinfo{person}{Ilya Sutskever}, {and}
  \bibinfo{person}{Dario Amodei}.} \bibinfo{year}{2020}\natexlab{}.
\newblock \showarticletitle{Language Models are Few-Shot Learners}. In \bibinfo{booktitle}{\emph{Advances in Neural Information Processing Systems}}, \bibfield{editor}{\bibinfo{person}{H.~Larochelle}, \bibinfo{person}{M.~Ranzato}, \bibinfo{person}{R.~Hadsell}, \bibinfo{person}{M.F. Balcan}, {and} \bibinfo{person}{H.~Lin}} (Eds.), Vol.~\bibinfo{volume}{33}. \bibinfo{publisher}{Curran Associates, Inc.}, \bibinfo{pages}{1877--1901}.
\newblock
\urldef\tempurl%
\url{https://proceedings.neurips.cc/paper_files/paper/2020/file/1457c0d6bfcb4967418bfb8ac142f64a-Paper.pdf}
\showURL{%
\tempurl}


\bibitem[Choquette et~al\mbox{.}(2021)]%
        {a100}
\bibfield{author}{\bibinfo{person}{Jack Choquette}, \bibinfo{person}{Wishwesh Gandhi}, \bibinfo{person}{Olivier Giroux}, \bibinfo{person}{Nick Stam}, {and} \bibinfo{person}{Ronny Krashinsky}.} \bibinfo{year}{2021}\natexlab{}.
\newblock \showarticletitle{NVIDIA A100 Tensor Core GPU: Performance and Innovation}.
\newblock \bibinfo{journal}{\emph{IEEE Micro}} \bibinfo{volume}{41}, \bibinfo{number}{2} (\bibinfo{year}{2021}), \bibinfo{pages}{29--35}.
\newblock
\urldef\tempurl%
\url{https://doi.org/10.1109/MM.2021.3061394}
\showDOI{\tempurl}


\bibitem[Chowdhery et~al\mbox{.}(2023)]%
        {PaLM}
\bibfield{author}{\bibinfo{person}{Aakanksha Chowdhery}, \bibinfo{person}{Sharan Narang}, \bibinfo{person}{Jacob Devlin}, \bibinfo{person}{Maarten Bosma}, \bibinfo{person}{Gaurav Mishra}, \bibinfo{person}{Adam Roberts}, \bibinfo{person}{Paul Barham}, \bibinfo{person}{Hyung~Won Chung}, \bibinfo{person}{Charles Sutton}, \bibinfo{person}{Sebastian Gehrmann}, \bibinfo{person}{Parker Schuh}, \bibinfo{person}{Kensen Shi}, \bibinfo{person}{Sashank Tsvyashchenko}, \bibinfo{person}{Joshua Maynez}, \bibinfo{person}{Abhishek Rao}, \bibinfo{person}{Parker Barnes}, \bibinfo{person}{Yi Tay}, \bibinfo{person}{Noam Shazeer}, \bibinfo{person}{Vinodkumar Prabhakaran}, \bibinfo{person}{Emily Reif}, \bibinfo{person}{Nan Du}, \bibinfo{person}{Ben Hutchinson}, \bibinfo{person}{Reiner Pope}, \bibinfo{person}{James Bradbury}, \bibinfo{person}{Jacob Austin}, \bibinfo{person}{Michael Isard}, \bibinfo{person}{Guy Gur-Ari}, \bibinfo{person}{Pengcheng Yin}, \bibinfo{person}{Toju Duke}, \bibinfo{person}{Anselm Levskaya},
  \bibinfo{person}{Sanjay Ghemawat}, \bibinfo{person}{Sunipa Dev}, \bibinfo{person}{Henryk Michalewski}, \bibinfo{person}{Xavier Garcia}, \bibinfo{person}{Vedant Misra}, \bibinfo{person}{Kevin Robinson}, \bibinfo{person}{Liam Fedus}, \bibinfo{person}{Denny Zhou}, \bibinfo{person}{Daphne Ippolito}, \bibinfo{person}{David Luan}, \bibinfo{person}{Hyeontaek Lim}, \bibinfo{person}{Barret Zoph}, \bibinfo{person}{Alexander Spiridonov}, \bibinfo{person}{Ryan Sepassi}, \bibinfo{person}{David Dohan}, \bibinfo{person}{Shivani Agrawal}, \bibinfo{person}{Mark Omernick}, \bibinfo{person}{Andrew~M. Dai}, \bibinfo{person}{Thanumalayan~Sankaranarayana Pillai}, \bibinfo{person}{Marie Pellat}, \bibinfo{person}{Aitor Lewkowycz}, \bibinfo{person}{Erica Moreira}, \bibinfo{person}{Rewon Child}, \bibinfo{person}{Oleksandr Polozov}, \bibinfo{person}{Katherine Lee}, \bibinfo{person}{Zongwei Zhou}, \bibinfo{person}{Xuezhi Wang}, \bibinfo{person}{Brennan Saeta}, \bibinfo{person}{Mark Diaz}, \bibinfo{person}{Orhan Firat},
  \bibinfo{person}{Michele Catasta}, \bibinfo{person}{Jason Wei}, \bibinfo{person}{Kathy Meier-Hellstern}, \bibinfo{person}{Douglas Eck}, \bibinfo{person}{Jeff Dean}, \bibinfo{person}{Slav Petrov}, {and} \bibinfo{person}{Noah Fiedel}.} \bibinfo{year}{2023}\natexlab{}.
\newblock \showarticletitle{PaLM: scaling language modeling with pathways}.
\newblock \bibinfo{journal}{\emph{J. Mach. Learn. Res.}} \bibinfo{volume}{24}, \bibinfo{number}{1}, Article \bibinfo{articleno}{240} (\bibinfo{date}{Jan.} \bibinfo{year}{2023}), \bibinfo{numpages}{113}~pages.
\newblock
\showISSN{1532-4435}


\bibitem[Deng et~al\mbox{.}(2019)]%
        {tie}
\bibfield{author}{\bibinfo{person}{Chunhua Deng}, \bibinfo{person}{Fangxuan Sun}, \bibinfo{person}{Xuehai Qian}, \bibinfo{person}{Jun Lin}, \bibinfo{person}{Zhongfeng Wang}, {and} \bibinfo{person}{Bo Yuan}.} \bibinfo{year}{2019}\natexlab{}.
\newblock \showarticletitle{TIE: energy-efficient tensor train-based inference engine for deep neural network}. In \bibinfo{booktitle}{\emph{Proceedings of the 46th International Symposium on Computer Architecture}} (Phoenix, Arizona) \emph{(\bibinfo{series}{ISCA '19})}. \bibinfo{publisher}{Association for Computing Machinery}, \bibinfo{address}{New York, NY, USA}, \bibinfo{pages}{264–278}.
\newblock
\showISBNx{9781450366694}
\urldef\tempurl%
\url{https://doi.org/10.1145/3307650.3322258}
\showDOI{\tempurl}


\bibitem[Frantar et~al\mbox{.}(2022)]%
        {gptq}
\bibfield{author}{\bibinfo{person}{Elias Frantar}, \bibinfo{person}{Saleh Ashkboos}, \bibinfo{person}{Torsten Hoefler}, {and} \bibinfo{person}{Dan Alistarh}.} \bibinfo{year}{2022}\natexlab{}.
\newblock \showarticletitle{{GPTQ}: Accurate Post-training Compression for Generative Pretrained Transformers}.
\newblock \bibinfo{journal}{\emph{arXiv preprint arXiv:2210.17323}} (\bibinfo{year}{2022}).
\newblock


\bibitem[Hajimolahoseini et~al\mbox{.}(2021)]%
        {compressing}
\bibfield{author}{\bibinfo{person}{Habib Hajimolahoseini}, \bibinfo{person}{Mehdi Rezagholizadeh}, \bibinfo{person}{Vahid Partovinia}, \bibinfo{person}{Marzieh Tahaei}, \bibinfo{person}{Omar~Mohamed Awad}, {and} \bibinfo{person}{Yang Liu}.} \bibinfo{year}{2021}\natexlab{}.
\newblock \showarticletitle{Compressing pre-trained language models using progressive low rank decomposition}.
\newblock \bibinfo{journal}{\emph{Advances in Neural Information Processing Systems}}  \bibinfo{volume}{35} (\bibinfo{year}{2021}), \bibinfo{pages}{6--14}.
\newblock


\bibitem[Han et~al\mbox{.}(2024)]%
        {robust_svd}
\bibfield{author}{\bibinfo{person}{Sangil Han}, \bibinfo{person}{Sungkyu Jung}, {and} \bibinfo{person}{Kyoowon Kim}.} \bibinfo{year}{2024}\natexlab{}.
\newblock \showarticletitle{Robust {SVD} Made Easy: A fast and reliable algorithm for large-scale data analysis}. In \bibinfo{booktitle}{\emph{Proceedings of The 27th International Conference on Artificial Intelligence and Statistics}} \emph{(\bibinfo{series}{Proceedings of Machine Learning Research}, Vol.~\bibinfo{volume}{238})}, \bibfield{editor}{\bibinfo{person}{Sanjoy Dasgupta}, \bibinfo{person}{Stephan Mandt}, {and} \bibinfo{person}{Yingzhen Li}} (Eds.). \bibinfo{publisher}{PMLR}, \bibinfo{pages}{1765--1773}.
\newblock
\urldef\tempurl%
\url{https://proceedings.mlr.press/v238/han24a.html}
\showURL{%
\tempurl}


\bibitem[Hinton et~al\mbox{.}(2015)]%
        {knowledge_distillation}
\bibfield{author}{\bibinfo{person}{Geoffrey Hinton}, \bibinfo{person}{Oriol Vinyals}, {and} \bibinfo{person}{Jeff Dean}.} \bibinfo{year}{2015}\natexlab{}.
\newblock \bibinfo{title}{Distilling the Knowledge in a Neural Network}.
\newblock
\newblock
\showeprint[arxiv]{1503.02531}~[stat.ML]
\urldef\tempurl%
\url{https://arxiv.org/abs/1503.02531}
\showURL{%
\tempurl}


\bibitem[Hrinchuk et~al\mbox{.}(2020)]%
        {tensorized}
\bibfield{author}{\bibinfo{person}{Oleksii Hrinchuk}, \bibinfo{person}{Valentin Khrulkov}, \bibinfo{person}{Leyla Mirvakhabova}, \bibinfo{person}{Elena Orlova}, {and} \bibinfo{person}{Ivan Oseledets}.} \bibinfo{year}{2020}\natexlab{}.
\newblock \showarticletitle{Tensorized Embedding Layers}. In \bibinfo{booktitle}{\emph{Findings of the Association for Computational Linguistics: EMNLP 2020}}, \bibfield{editor}{\bibinfo{person}{Trevor Cohn}, \bibinfo{person}{Yulan He}, {and} \bibinfo{person}{Yang Liu}} (Eds.). \bibinfo{publisher}{Association for Computational Linguistics}, \bibinfo{address}{Online}, \bibinfo{pages}{4847--4860}.
\newblock
\urldef\tempurl%
\url{https://doi.org/10.18653/v1/2020.findings-emnlp.436}
\showDOI{\tempurl}


\bibitem[Hu et~al\mbox{.}(2021)]%
        {LoRA}
\bibfield{author}{\bibinfo{person}{Edward~J. Hu}, \bibinfo{person}{Yelong Shen}, \bibinfo{person}{Phillip Wallis}, \bibinfo{person}{Zeyuan Allen{-}Zhu}, \bibinfo{person}{Yuanzhi Li}, \bibinfo{person}{Shean Wang}, {and} \bibinfo{person}{Weizhu Chen}.} \bibinfo{year}{2021}\natexlab{}.
\newblock \showarticletitle{LoRA: Low-Rank Adaptation of Large Language Models}.
\newblock \bibinfo{journal}{\emph{CoRR}}  \bibinfo{volume}{abs/2106.09685} (\bibinfo{year}{2021}).
\newblock
\showeprint[arXiv]{2106.09685}
\urldef\tempurl%
\url{https://arxiv.org/abs/2106.09685}
\showURL{%
\tempurl}


\bibitem[Kopiczko et~al\mbox{.}(2023)]%
        {Kopiczko}
\bibfield{author}{\bibinfo{person}{Dawid~J Kopiczko}, \bibinfo{person}{Tijmen Blankevoort}, {and} \bibinfo{person}{Yuki~M Asano}.} \bibinfo{year}{2023}\natexlab{}.
\newblock \showarticletitle{{VeRA}: Vector-based Random Matrix Adaptation}.
\newblock \bibinfo{journal}{\emph{arXiv [cs.CL]}} (\bibinfo{date}{Oct.} \bibinfo{year}{2023}).
\newblock


\bibitem[Lin et~al\mbox{.}(2019)]%
        {holistic}
\bibfield{author}{\bibinfo{person}{Shaohui Lin}, \bibinfo{person}{Rongrong Ji}, \bibinfo{person}{Chao Chen}, \bibinfo{person}{Dacheng Tao}, {and} \bibinfo{person}{Jiebo Luo}.} \bibinfo{year}{2019}\natexlab{}.
\newblock \showarticletitle{Holistic CNN Compression via Low-Rank Decomposition with Knowledge Transfer}.
\newblock \bibinfo{journal}{\emph{IEEE Transactions on Pattern Analysis and Machine Intelligence}} \bibinfo{volume}{41}, \bibinfo{number}{12} (\bibinfo{year}{2019}), \bibinfo{pages}{2889--2905}.
\newblock
\urldef\tempurl%
\url{https://doi.org/10.1109/TPAMI.2018.2873305}
\showDOI{\tempurl}


\bibitem[Martins et~al\mbox{.}(2015)]%
        {martins2015open}
\bibfield{author}{\bibinfo{person}{Mayler Martins}, \bibinfo{person}{Jody~Maick Matos}, \bibinfo{person}{Renato~P Ribas}, \bibinfo{person}{Andr{\'e} Reis}, \bibinfo{person}{Guilherme Schlinker}, \bibinfo{person}{Lucio Rech}, {and} \bibinfo{person}{Jens Michelsen}.} \bibinfo{year}{2015}\natexlab{}.
\newblock \showarticletitle{Open cell library in 15nm freepdk technology}. In \bibinfo{booktitle}{\emph{Proceedings of the 2015 Symposium on International Symposium on Physical Design}}. \bibinfo{pages}{171--178}.
\newblock


\bibitem[Moar et~al\mbox{.}(2024)]%
        {IISWC_paper}
\bibfield{author}{\bibinfo{person}{Chakshu Moar}, \bibinfo{person}{Faraz Tahmasebi}, \bibinfo{person}{Michael Pellauer}, {and} \bibinfo{person}{Hyoukjun Kwon}.} \bibinfo{year}{2024}\natexlab{}.
\newblock \showarticletitle{Characterizing the Accuracy-Efficiency Trade-off of Low-rank Decomposition in Language Models}. In \bibinfo{booktitle}{\emph{2024 IEEE International Symposium on Workload Characterization (IISWC)}}. \bibinfo{pages}{194--209}.
\newblock
\urldef\tempurl%
\url{https://doi.org/10.1109/IISWC63097.2024.00026}
\showDOI{\tempurl}


\bibitem[NVIDIA(2023)]%
        {nvidia_h100}
\bibfield{author}{\bibinfo{person}{NVIDIA}.} \bibinfo{year}{2023}\natexlab{}.
\newblock \bibinfo{title}{NVIDIA H100 Tensor Core GPU}.
\newblock \bibinfo{howpublished}{\url{https://www.nvidia.com/en-us/data-center/h100/}}.
\newblock


\bibitem[OpenAI(2023)]%
        {GPT-4}
\bibfield{author}{\bibinfo{person}{OpenAI}.} \bibinfo{year}{2023}\natexlab{}.
\newblock \showarticletitle{{GPT-4} Technical Report}.
\newblock \bibinfo{journal}{\emph{CoRR}}  \bibinfo{volume}{abs/2303.08774} (\bibinfo{year}{2023}).
\newblock
\urldef\tempurl%
\url{https://doi.org/10.48550/ARXIV.2303.08774}
\showDOI{\tempurl}
\showeprint[arXiv]{2303.08774}


\bibitem[Phan et~al\mbox{.}(2020)]%
        {stable}
\bibfield{author}{\bibinfo{person}{Anh-Huy Phan}, \bibinfo{person}{Konstantin Sobolev}, \bibinfo{person}{Konstantin Sozykin}, \bibinfo{person}{Dmitry Ermilov}, \bibinfo{person}{Julia Gusak}, \bibinfo{person}{Petr Tichavsky}, \bibinfo{person}{Valeriy Glukhov}, \bibinfo{person}{Ivan Oseledets}, {and} \bibinfo{person}{Andrzej Cichocki}.} \bibinfo{year}{2020}\natexlab{}.
\newblock \bibinfo{title}{Stable Low-rank Tensor Decomposition for Compression of Convolutional Neural Network}.
\newblock
\newblock
\showeprint[arxiv]{2008.05441}~[cs.CV]
\urldef\tempurl%
\url{https://arxiv.org/abs/2008.05441}
\showURL{%
\tempurl}


\bibitem[Saha et~al\mbox{.}(2024)]%
        {low_prec_low_rank}
\bibfield{author}{\bibinfo{person}{Rajarshi Saha}, \bibinfo{person}{Naomi Sagan}, \bibinfo{person}{Varun Srivastava}, \bibinfo{person}{Andrea~J. Goldsmith}, {and} \bibinfo{person}{Mert Pilanci}.} \bibinfo{year}{2024}\natexlab{}.
\newblock \bibinfo{title}{Compressing Large Language Models using Low Rank and Low Precision Decomposition}.
\newblock
\newblock
\showeprint[arxiv]{2405.18886}~[cs.LG]
\urldef\tempurl%
\url{https://arxiv.org/abs/2405.18886}
\showURL{%
\tempurl}


\bibitem[Sun et~al\mbox{.}(2024)]%
        {pruning}
\bibfield{author}{\bibinfo{person}{Mingjie Sun}, \bibinfo{person}{Zhuang Liu}, \bibinfo{person}{Anna Bair}, {and} \bibinfo{person}{J.~Zico Kolter}.} \bibinfo{year}{2024}\natexlab{}.
\newblock \showarticletitle{A Simple and Effective Pruning Approach for Large Language Models}. In \bibinfo{booktitle}{\emph{The Twelfth International Conference on Learning Representations, {ICLR} 2024, Vienna, Austria, May 7-11, 2024}}. \bibinfo{publisher}{OpenReview.net}.
\newblock
\urldef\tempurl%
\url{https://openreview.net/forum?id=PxoFut3dWW}
\showURL{%
\tempurl}


\bibitem[Touvron et~al\mbox{.}(2023)]%
        {llama-2-7b}
\bibfield{author}{\bibinfo{person}{Hugo Touvron}, \bibinfo{person}{Louis Martin}, \bibinfo{person}{Kevin~R. Stone}, \bibinfo{person}{Peter Albert}, \bibinfo{person}{Amjad Almahairi}, \bibinfo{person}{Yasmine Babaei}, \bibinfo{person}{Nikolay Bashlykov}, \bibinfo{person}{Soumya Batra}, \bibinfo{person}{Prajjwal Bhargava}, \bibinfo{person}{Shruti Bhosale}, \bibinfo{person}{Dan Bikel}, \bibinfo{person}{Lukas Blecher}, \bibinfo{person}{Cristian Canton~Ferrer}, \bibinfo{person}{Moya Chen}, \bibinfo{person}{Guillem Cucurull}, \bibinfo{person}{David Esiobu}, \bibinfo{person}{Jude Fernandes}, \bibinfo{person}{Jeremy Fu}, \bibinfo{person}{Wenyin Fu}, \bibinfo{person}{Brian Fuller}, \bibinfo{person}{Cynthia Gao}, \bibinfo{person}{Vedanuj Goswami}, \bibinfo{person}{Naman Goyal}, \bibinfo{person}{Anthony Hartshorn}, \bibinfo{person}{Saghar Hosseini}, \bibinfo{person}{Rui Hou}, \bibinfo{person}{Hakan Inan}, \bibinfo{person}{Marcin Kardas}, \bibinfo{person}{Viktor Kerkez}, \bibinfo{person}{Madian Khabsa},
  \bibinfo{person}{Isabel~M. Kloumann}, \bibinfo{person}{Artem Korenev}, \bibinfo{person}{Punit~Singh Koura}, \bibinfo{person}{Marie‑Anne Lachaux}, \bibinfo{person}{Thibaut Lavril}, \bibinfo{person}{Jenya Lee}, \bibinfo{person}{Diana Liskovich}, \bibinfo{person}{Yinghai Lu}, \bibinfo{person}{Yuning Mao}, \bibinfo{person}{Xavier Martinet}, \bibinfo{person}{Todor Mihaylov}, \bibinfo{person}{Pushkar Mishra}, \bibinfo{person}{Igor Molybog}, \bibinfo{person}{Yixin Nie}, \bibinfo{person}{Andrew Poulton}, \bibinfo{person}{Jeremy Reizenstein}, \bibinfo{person}{Rashi Rungta}, \bibinfo{person}{Kalyan Saladi}, \bibinfo{person}{Alan Schelten}, \bibinfo{person}{Ruan Silva}, \bibinfo{person}{Eric~Michael Smith}, \bibinfo{person}{Ranjan Subramanian}, \bibinfo{person}{Xia Tan}, \bibinfo{person}{Binh Tang}, \bibinfo{person}{Ross Taylor}, \bibinfo{person}{Adina Williams}, \bibinfo{person}{Jian~Xiang Kuan}, \bibinfo{person}{Puxin Xu}, \bibinfo{person}{Zhengxu Yan}, \bibinfo{person}{Iliyan Zarov}, \bibinfo{person}{Yuchen
  Zhang}, \bibinfo{person}{Angela Fan}, \bibinfo{person}{Melanie Kambadur}, \bibinfo{person}{Sharan Narang}, \bibinfo{person}{Aurelien Rodriguez}, \bibinfo{person}{Robert Stojnic}, \bibinfo{person}{Sergey Edunov}, {and} \bibinfo{person}{Thomas Scialom}.} \bibinfo{year}{2023}\natexlab{}.
\newblock \showarticletitle{Llama-2: Open Foundation and Fine‑Tuned Chat Models}.
\newblock \bibinfo{journal}{\emph{arXiv preprint arXiv:2307.09288}} (\bibinfo{year}{2023}).
\newblock
\urldef\tempurl%
\url{https://doi.org/10.48550/arXiv.2307.09288}
\showDOI{\tempurl}


\bibitem[Xu et~al\mbox{.}(2012)]%
        {robust_pca}
\bibfield{author}{\bibinfo{person}{Huan Xu}, \bibinfo{person}{Constantine Caramanis}, {and} \bibinfo{person}{Sujay Sanghavi}.} \bibinfo{year}{2012}\natexlab{}.
\newblock \showarticletitle{Robust PCA via Outlier Pursuit}.
\newblock \bibinfo{journal}{\emph{IEEE Transactions on Information Theory}} \bibinfo{volume}{58}, \bibinfo{number}{5} (\bibinfo{year}{2012}), \bibinfo{pages}{3047--3064}.
\newblock
\urldef\tempurl%
\url{https://doi.org/10.1109/TIT.2011.2173156}
\showDOI{\tempurl}


\bibitem[Zhang et~al\mbox{.}(2023)]%
        {adalora}
\bibfield{author}{\bibinfo{person}{Qingru Zhang}, \bibinfo{person}{Minshuo Chen}, \bibinfo{person}{Alexander Bukharin}, \bibinfo{person}{Nikos Karampatziakis}, \bibinfo{person}{Pengcheng He}, \bibinfo{person}{Yu Cheng}, \bibinfo{person}{Weizhu Chen}, {and} \bibinfo{person}{Tuo Zhao}.} \bibinfo{year}{2023}\natexlab{}.
\newblock \bibinfo{title}{AdaLoRA: Adaptive Budget Allocation for Parameter-Efficient Fine-Tuning}.
\newblock
\newblock
\showeprint[arxiv]{2303.10512}~[cs.CL]
\urldef\tempurl%
\url{https://arxiv.org/abs/2303.10512}
\showURL{%
\tempurl}


\end{thebibliography}
%%%%%%%%%%%%%%%%%%%%%%%%%%%%%%%%%%%%

\end{document}